\documentclass[twocolumn,floatfix,superscriptaddress,amsmath,showpacs,showkeys,aps,prb,comment]{revtex4}

\usepackage[final]{graphicx}
\usepackage{t1enc}
\usepackage{bm}

\begin{document}

\bibliographystyle{apsrev}

\title{\bf Comment on ``Phase transitions in a square Ising model with exchange and dipole interactions''
by E. Rastelli, S. Regina and A. Tassi}

\author{Sergio A. Cannas}
\email{cannas@famaf.unc.edu.ar} \affiliation{Facultad de
Matem\'atica, Astronom\'{\i}a y F\'{\i}sica, Universidad Nacional
de C\'ordoba, \\ Ciudad Universitaria, 5000 C\'ordoba, Argentina}
\altaffiliation{Member of CONICET, Argentina}
\author{Mateus F. Michelon}
\email{michelon@ifi.unicamp.br} \affiliation{Departamento de F\'{\i}sica,
Universidade Federal do Rio Grande do Sul\\
CP 15051, 91501--979, Porto Alegre, Brazil}
\altaffiliation{Present address: Instituto de Física Gleb Whataghin, UNICAMP, Campinas, Brazil}
\author{Daniel A. Stariolo}
\email{stariolo@if.ufrgs.br}
\affiliation{Departamento de F\'{\i}sica,
Universidade Federal do Rio Grande do Sul\\
CP 15051, 91501--979, Porto Alegre, Brazil}
\altaffiliation{Research Associate of the Abdus Salam International Centre for Theoretical Physics,
Trieste, Italy}
\author{Francisco A. Tamarit}
\email{tamarit@famaf.unc.edu.ar} \affiliation{Facultad de  Matem\'atica, Astronom\'{\i}a
y F\'{\i}sica, Universidad Nacional de C\'ordoba, \\ Ciudad Universitaria, 5000
C\'ordoba, Argentina} \altaffiliation{Member of CONICET, Argentina}

\date{\today}

\begin{abstract}

\end{abstract}


\maketitle

In an recent paper\cite{RaReTa2006} Regina, Rastelli and Tassi
investigated the critical properties of a two dimensional Ising
model with exchange and dipolar interactions described by the
Hamiltonian

\begin{equation}
{\cal H}= - J \sum_{<i,j>} \sigma_i \sigma_j + g  \sum_{i\neq j}
\frac{\sigma_i \sigma_j}{r^3_{ij}} \label{Hamilton1}
\end{equation}

\noindent for different values of the ratio $J/g$ using Monte
Carlo (MC) simulations and finite size scaling for different
system sizes. Hamiltonian (\ref{Hamilton1}) describes an ultrathin
metal-on-metal magnetic film with perpendicular anisotropy in the
monolayer limit\cite{DeMaWh2000}. Almost simultaneously we
published a paper\cite{CaMiStTa2006} about the same subject with a
very similar analysis for a different set of values of the ratio
$J/g$.  Our detailed numerical analysis showed evidences of an
intermediate disordered phase, which was not detected by Rastelli
analysis. The particular characteristics of that intermediate
phase and its associated phase transitions can introduce a strong
bias in the results if not properly taken into account by the
numerical procedure. In this comment we show that the numerical
calculation protocol used by Rastelli et al is unable to detect
the presence of the intermediate phase leading to spurious
results. Therefore some of their conclusions must be revised.

The main point concerns the results of Rastelli et al for
$J/g=3.4$ and $J/g=5$ compared with our results in the same region
of the phase diagram, namely for $J/g=4$ ($\delta=J/(2\, g)=2$ in
our notation; temperatures in our paper are also rescaled by a
factor 2 respect to those in Rastelli et al due to a different
choice of the energy scales; see Ref.\cite{CaMiStTa2006} for
details). For that values of $J/g$ Rastelli et all found a single
peak in the specific heat and single minimum in the fourth order
cumulant, whose finite size scaling is consistent with a first
order phase transition, concluding that the system presents a
unique first order phase transition between the (low temperature)
striped and the (high temperature) tetragonal liquid phases. On
the other hand, we showed in Ref.\cite{CaMiStTa2006} that for
$\delta=2$ the specific heat presents two distinct maxima and the
fourth order cumulant two distinct minima, consistent with the
existence of an intermediate phase, characterized by orientational
order and positional disorder, evidenced in the finite size
scaling behavior of the static structure factor. The existence of
an intermediate, {\it nematic phase}, is consistent with one of
the two possible scenarios predicted by a continuum approximation
for ultrathin magnetic films in
Refs.\cite{KaPo1993b,AbKaPoSa1995}.
\begin{figure}
\begin{center}
\includegraphics[scale=0.6,angle=0]{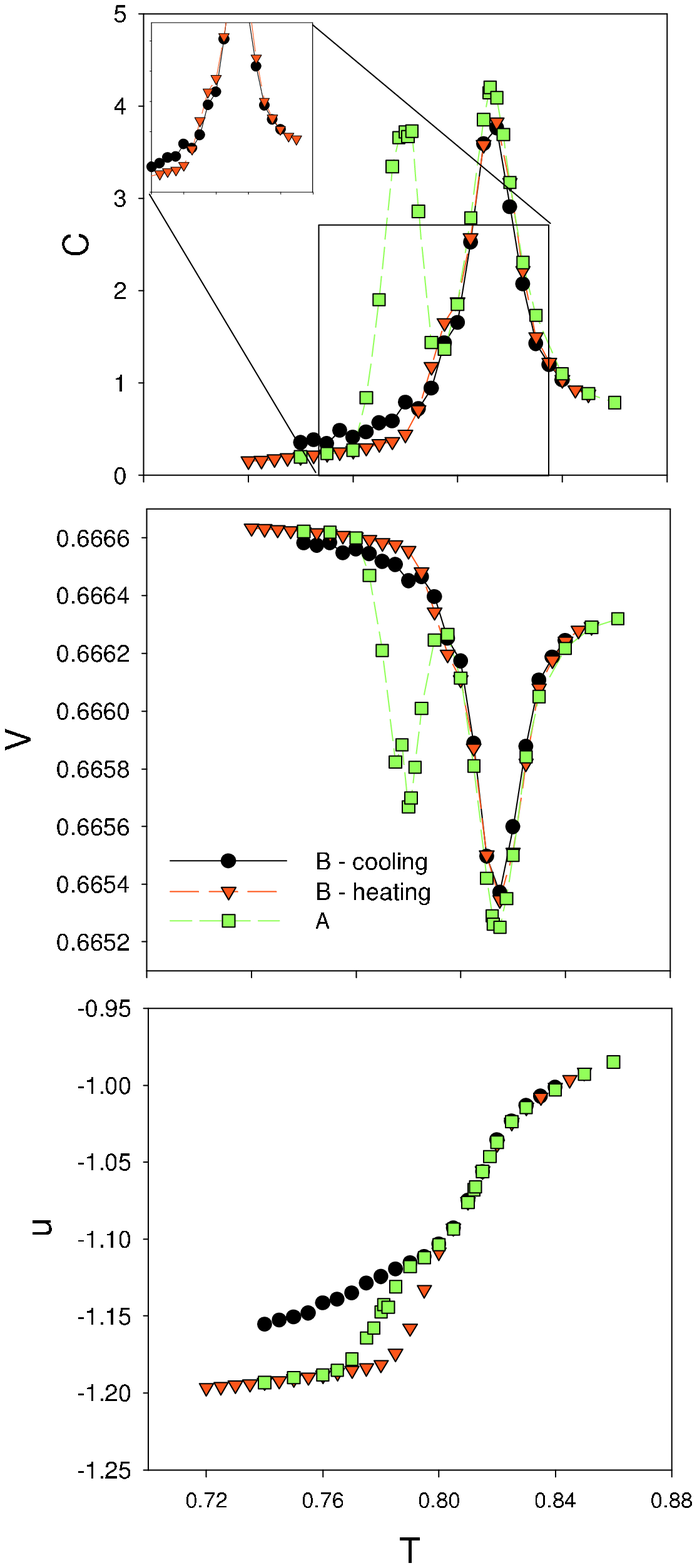}
\caption{\label{fig1} (Color on-line) Specific heat $C$, fourth
order cumulant $V$ and internal energy per spin $u$ as a function
of temperature for $J/g=4$ ($\delta=2$) and $L=48$, obtained from
different MC simulation protocols (see text for details).}
\end{center}
\end{figure}
Moreover, we showed that the finite size scaling of the different
thermodynamical quantities around the nematic-tetragonal phase
transition are consistent with a first order transition, while
those observed around the stripe-nematic transition present
unusual features, some of them resembling a Kosterlitz-Thouless
one (like a saturation in the specific heat maximum and a
continuous change in the internal energy) while others are also
characteristic of a first order phase transition. In particular,
one of those characteristics is the existence of diverging free
energy barriers between the nematic and the stripe phases, which
generate a strong meta--stabilities  when the system is cooled or
heated through the transition temperature. To correctly
characterize the finite size scaling at this phase transition
thermodynamical quantities must be averaged for every temperature
over a large single MC run, thus allowing the system to sample
properly  the different phases. Otherwise, as explained by Challa
et al\cite{ChLaBi1986},  the first-order character of the
transition introduces pronounced meta--stabilities with the system
spending most of the time in one of the two phases (see
Ref.\cite{ChLaBi1986} for details). Hence, for every temperature
we first let the system run over a transient period of $M_1$ Monte
Carlo Steps (MCS) and then we averaged the different quantities
over a single MC run of $M_2$ MCS, with $M_1$ ranging between
$10^6$ and $2 \times 10^7$, and $M_2$ ranging between $2 \times
10^7$  and $2 \times 10^8$; we shall call this procedure protocol
A. We believe that the absence of the nematic-stripe phase
transition (and therefore the evidence of the nematic phase) in
Rastelli et al paper is due to the particular MC protocol they
used, which introduced a strong bias in their results. They took
averages over eight independent MC runs of $10^5$ MCS for each
temperature, taking the initial configuration for every
temperature as the final configuration of the previous one,
disregarding $10^4$ MCS for thermalization\cite{RaReTa2006}; we
shall call this procedure protocol B. This protocol is similar to
a finite  rate heating (cooling) procedure from low (high)
temperature and the system can get trapped in the meta--stable
striped (nematic) state, thus hiding the low temperature phase
transition, whose associated free energy barriers are very high
for system sizes used in that work. Indeed we checked this
assumption by computing the moments of the energy, namely the
internal energy per spin $u$, the specific heat $C$ and the fourth
order cumulant $V$ as a function of the temperature (see
definitions in Ref.\cite{CaMiStTa2006}) using Rastelli et al MC
protocol for $\delta=2$ and $L=48$, both heating from the low
temperature stable phase and cooling from the high temperature
one. The results were averaged over  $8$ (as in Rastelli et al
paper) and $50$ independent runs; no qualitative differences were
observed. The obtained results for $50$ independent runs are
 compared in Fig.\ref{fig1} with the results obtained in Ref.\cite{CaMiStTa2006} for the same
parameters values.

We see that the low temperature maximum of $C$  and the minimum of
$V$ are almost undetectable by Rastelli et al MC protocol, the
only noticeable effect being some enlarged fluctuations in the
cooling procedure and a very small shoulder for the heating result
at the left of the high temperature specific heat maximum (the
inset of Fig.\ref{fig1} remarks this effect) or cumulant minimum,
but clearly separated from the actual low temperature transition.
Moreover, the behavior of $u$ shows clearly that the system gets
trapped in a metastable super-cooled (super-heated) state, as in a
finite rate cooling (heating) process (compare with Fig.19 in
Ref.\cite{CaMiStTa2006}). Indeed we observed that for some
individual runs in the heating procedure the specific heat
presents a small secondary maximum of the specific heat at low
temperatures, but those maxima are a result of a spinodal
instability of the super-heated state and therefore are strongly
shifted to the right of the actual maximum. Moreover, those
spinodal maxima appear located at different random temperatures
and they are smeared out when averaged, giving rise to the
observed shoulder.

Even when the analysis in Ref.\cite{CaMiStTa2006} was carried out
for the particular value $\delta=2$, theoretical
results\cite{KaPo1993b,AbKaPoSa1995} lead us  to expect the
presence of the nematic phase to be found for a wide range of
values of $\delta$, although possibly for even narrower ranges of
temperatures than for $\delta=2$. In that case, the above
described spurious effects should become worse. Indeed, the
nematic phase and the associated behavior in the specific heat and
cumulant has been found\cite{PiCa2006} using protocol A for other
values of $\delta$, including $\delta=2.25$ ($J/g=4.5$), which
corresponds to the $h=3$ ground state region, as the $J/g=5$
 case analyzed by Rastelli et al (and close to it).

 Another related point concerns the scaling of the
principal maximum of $C$ for $J/g=5$ in Rastelli et al work. They
found that the maximum scales as $~ L^{1.2}$ instead of the $L^2$
scaling expected in a first order transition. A careful analysis
of the orientational order parameter histogram for $\delta=2$
shows the presence of multiple peaks around this transition (See
Fig.7 in Ref.\cite{CaMiStTa2006}) indicating  the presence of a
complex multiple phase structure in a very narrow range of
temperatures; such structure is only detectable for large system
sizes and can introduce a strong bias in the scaling properties
for small sizes. The anomalous scaling observed by Rastelli et all
is probably due to this finite size effect.

The above considerations also apply  to the case $J/g=6$. In this
case the results of Rastelli et al (see Fig.12 in
Ref.\cite{RaReTa2006}) show rather clearly the presence of a
secondary low temperature peak in the specific heat, whose height
is independent of the system size, very similar to that observed
for $J/g=4$ (see Fig.9 in Ref.\cite{CaMiStTa2006}). There is also
some indication of a vanishing secondary minimum in $V$, similar
to that observed for $J/g=4$. Although those anomalies are very
small, they are probably affected  also by the simulation protocol
and by finite size effects, which in this system become stronger
as $J/g$ increases. Based on the scaling of the specific heat
maximum and on the apparently vanishing minimum of the energy
cumulant the authors conclude that the transition is continuous
for $J/g \geq 6$. However, such results can also be consistent
with a weak first order transition that continuously fades out as
$J/g$ increases. If that would be the case, detecting the order of
the transition becomes more delicate due to the above described
finite size effects. Indeed, a Hartree approximation applied to
the continuous version of Hamiltonian (\ref{Hamilton1}) predicts a
first order transition for a wide range of values of
\cite{CaStTa2004} $J/g$. Rastelli et al criticize this result
claiming that such approximation wrongly predicts a first order
transition also for $g=0$, where it is known that it is second
order.
As is well known since the work by Brazovskii~\cite{Br1975}, that
calculation breaks down for $g=0$. Nevertheless, for a wide range
of values of $J/g > 1$ the approximation gives excellent qualitative
results. It is at least very suggestive that our numerical results
seem to be in agreement with the predictions of this theory.
To be more specific, at the transition a modulated solution of the form
$\phi(\vec{r})=m\cos{(\vec{k_0}\cdot \vec{r})}$ becomes stable and
the amplitude of the equilibrium solution $m(T)$ jumps
discontinously to a non zero value. Following
Brazovskii~\cite{Br1975} it is possible to show that the amplitude
at the transition is given by
\begin{equation}
m_c=m(T_c) \propto \sqrt{r_m(T_c)/u},
\end{equation}
where $u$ is the coupling constant
of the Landau energy and $r_m$ is the renormalized mass of the
modulated solution~\cite{CaStTa2004}. At the transition point $r_m \propto (u\,k_0^2)^{2/3}$
and
\begin{equation}
m_c \propto \frac{k_0^{2/3}}{u^{1/6}}
\end{equation}
In our model $k_0 \propto g/J$ and consequently the amplitude
at the transition decreases continuously as $g/J$ gets smaller within
the range of validity of the approximation. This
shows that the transition becomes weaker as $g/J$ diminishes.
This is completely consistent with the difficulty in the numerical
simulations to detect the first order nature of the transition for
small values of $g/J$.

Finally, for $J/g=1.7$, Rastelli et al found  evidence of a
continuous phase transition with unusual critical exponents, from
which they conclude that the $h=1$ spin configurations ($h$ being
the  width of the ground state stripes) undergo a continuous phase
transition with a different universality class from the case
$J=0$. In Ref.\cite{CaMiStTa2006} we presented evidence that the
system undergo a unique weak first order transition  for $J/g=2$
($\delta=1$); this value of the ratio also corresponds to the
$h=1$ region of the phase diagram (see phase diagram in
\cite{GlTaCaMo2003}). These results are consistent with the
presence of a second order transition line for small values of
$J/g$ that joins with continuous slope a first order transition
line for larger values of $J/g$ at a tricritical point in that
region of the phase diagram (see phase diagram in
\cite{GlTaCaMo2003}). Hence, the unusual critical exponents
observed by Rastelli et al are most probably due to a crossover
effect related to the presence of that tricritical point in the
neighborhood of $J/g=1.7$ and not to a new universality class
transition line.

 This work was partially supported by grants from CONICET
(Argentina), Agencia C\'ordoba Ciencia (Argentina), SeCyT,
Universidad Nacional de C\'ordoba (Argentina), CNPq (Brazil) and
ICTP grant NET-61 (Italy).

\end{document}